\documentclass[twocolumn,showpacs,showkeys]{revtex4}
\usepackage{graphicx}
\usepackage{bm}
\usepackage{color}
\usepackage{amsmath}
\usepackage{natbib}

\begin{document}

\title{Separated spin-up and spin-down quantum hydrodynamics of degenerated electrons: spin-electron acoustic wave appearance}

\author{Pavel A. Andreev}
\email{andreevpa@physics.msu.ru}
\affiliation{Faculty of physics, Lomonosov Moscow State University, Moscow, Russian Federation.}

 \date{\today}

\begin{abstract}
Quantum hydrodynamic (QHD) model of charged spin-1/2 particles contains physical quantities defined for all particles of a species including particles with spin-up and with spin-down. Different population of states with different spin direction is included in the spin density (magnetization). In this paper we derive a QHD model, which separately describes spin-up electrons and spin-down electrons. Hence we consider electrons with different projection of spin on the preferable direction as two different species of particles. We show that numbers of particles with different spin direction do not conserve. Hence the continuity equations contain sources of particles. These sources are caused by the interactions of spins with magnetic field. Terms of similar nature arise in the Euler equation. We have that z-projection of the spin density is no longer an independent variable. It is proportional to difference between concentrations of electrons with spin-up and electrons with spin-down. In terms of new model we consider propagation of waves in magnetized plasmas of degenerate electrons and motionless ions. We show that new form of QHD equations gives all solutions obtained from traditional form of QHD equations with no distinguish of spin-up and spin-down states. But it also reveals a sound-like solution we call the spin-electron acoustic wave. Coincidence of most solutions is expected since we started derivation with the same basic equation.
\end{abstract}

\pacs{52.30.Ex}% PACS, the Physics and Astronomy
                             % Classification Scheme.
\keywords{quantum plasmas, quantum hydrodynamics}
%Use showkeys class option if keyword

\maketitle

%52.30.Ex	Two-fluid and multi-fluid plasmas
%52.35.Dm	Sound waves

%52.27.Ep	Electron-positron plasmas

%%%%%%%%%%TEXT

\section{\label{sec:level1} Introduction}

Considering quantum plasmas of spinning particles we apply equations of the quantum hydrodynamics (QHD) or the quantum kinetics. Different methods of derivation of QHD equations were presented in Refs. \cite{MaksimovTMP 1999}, \cite{Haas PRE 00}, \cite{Manfredi PRB 01}, \cite{Koide PRC 13}, they have also been applied to quantum plasmas of spinning particles \cite{Koide PRC 13}, \cite{Maksimov Izv 2000}, \cite{MaksimovTMP 2001}. These equations contain the particle concentration $n(\textbf{r},t)$, the momentum density $\textbf{j}(\textbf{r},t)$, the velocity field $\textbf{v}(\textbf{r},t)$, the distribution function $f(\textbf{r},\textbf{p},t)$ describing all particles of a species independently of their spin direction. Difference between numbers of particles in different spin states is included in the spin density $\textbf{S}(\textbf{r},t)$ or magnetization $\textbf{M}(\textbf{r},t)=\gamma \textbf{S}(\textbf{r},t)$, where $\gamma$ is the gyromagnetic ratio. These models do not contain explicit distinguish between spin-up and spin-down states of particles.

Basic equations of many-particle quantum hydrodynamic of spin-1/2 particles were developed in 2000-2001 in Refs. \cite{Maksimov Izv 2000}, \cite{MaksimovTMP 2001} and \cite{MaksimovTMP 2001 b}. Further development of the method can be found in Refs. \cite{Andreev RPJ 07}, \cite{Andreev AtPhys 08}, \cite{Andreev IJMP 12}, \cite{Andreev spin current}, \cite{Andreev 1403 exchange} and \cite{Andreev PRB 11}. It includes explicit consideration of the spin-current \cite{Andreev RPJ 07} and spin-orbit \cite{Andreev IJMP 12} interactions. Derivation of the energy evolution equation \cite{MaksimovTMP 1999}, \cite{MaksimovTMP 2001}, \cite{Andreev RPJ 07} and the spin current (magnetization flux) evolution equation \cite{Andreev spin current} were performed. The exchange interaction was considered in Refs. \cite{MaksimovTMP 1999}, \cite{MaksimovTMP 2001 b}, \cite{Andreev 1403 exchange}. The QHD model for particles with electric dipole moment was developed in \cite{Andreev PRB 11}. All these developments were performed in terms of one method: method of many-particle quantum hydrodynamics suggested in Refs. \cite{MaksimovTMP 1999}, \cite{MaksimovTMP 2001}. Comprehensive analysis of quantum hydrodynamic equation for a single spin-1/2 particle in an external field had been performed by Takabayasi \cite{Takabayasi PTP 55 b}-\cite{Takabayasi PTP 83} during 50s of the twentieth century.

In the single fluid model of electrons with different spins the dynamic of spins is governed by the generalization of Bloch equation \cite{Takabayasi PTP 55 b}
\begin{equation}\label{SUSD eq of magnetic moments evol} n(\partial_{t}+\textbf{v}\nabla) \mbox{\boldmath $\mu$}-\frac{\hbar}{2m\gamma}\partial^{\beta}[n
\mbox{\boldmath $\mu$},\partial^{\beta}\mbox{\boldmath $\mu$}]=\frac{2\gamma}{\hbar}n[\mbox{\boldmath $\mu$},\textbf{B}],\end{equation}
where $\mbox{\boldmath $\mu$}$ is the reduced magnetization $\textbf{M}(\textbf{r},t)=n\mbox{\boldmath $\mu$}$, $[\textbf{a},\textbf{b}]$ is the vector product of vectors $\textbf{a}$ and $\textbf{b}$. The first groups of terms in the left-hand side of equation (\ref{SUSD eq of magnetic moments evol}) are the substantial derivative of the reduced magnetization. The second terms are the quantum Bohm potential for the Bloch equation. In the right-hand side of equation (\ref{SUSD eq of magnetic moments evol}) we have the torque caused by the interaction with the external magnetic field and the interparticle interactions. In two fluid model z-projection of magnetization $M_{z}$ is no longer an independent variable. It is proportional to difference of concentrations with spin up and spin-down. Other projections of the magnetization $M_{x}$ and $M_{y}$ appear in two fluid model as independent variables, but they do not wear indexes "up" or "down" being related to both species of electrons. It happens because definitions of $M_{x}$ and $M_{y}$ contain wave functions of spin-up and spin-down electrons.

Due to development of the field of quantum plasma \cite{Shukla PhUsp 2010}, \cite{Shukla RMP 11}, it is interesting to derive  a set of QHD equations of degenerate electrons considering two different spin states (spin-up and spin-down) as two different species of particles. In this paper we perform derivation of QHD equation explicitly distinguishing spin-up and spin-down states.

Different linear and non-linear excitations were considered in quantum plasmas \cite{Shukla PhUsp 2010}, \cite{Shukla RMP 11}. In this paper we also focus our attention on linear excitations in magnetized quantum plasmas of degenerate electrons and motionless ions in terms of new form of the QHD model. Basic linear phenomenon in spin-1/2 quantum plasmas were considered in Refs. \cite{Andreev AtPhys 08}, \cite{Andreev IJMP 12}, \cite{Marklund PRL07}, \cite{Brodin NJP 07}, \cite{Andreev VestnMSU 2007}, \cite{Misra JPP 10}, where contribution of spin in the dispersion of plasma waves was found and existence of spin-plasma waves was demonstrated. Electrons were considered as single fluid in these papers. We are going to find out that changes ay application of spin separated QHD.

This paper is organized as follows.  In Sec. II we derive the QHD model considering spin-up electrons and spin-down electrons as different species.  In Sec. III we consider propagation of waves parallel to external field as an illustration of derived equations. This problem has been solved in literature in term of usual QHD. We compare results of two different methods of fluidizations of the Pauli equation. In Sec. IV brief summary of obtained results is presented.

\section{\label{sec:level1} Model}

In this section we are going to derive the set of QHD equations for degenerate electrons considering spin-up and spin-down states as two different species. This derivation can be performed in terms of many-particle quantum hydrodynamics \cite{Maksimov Izv 2000}-\cite{Andreev PRB 11}. However, for simplicity of presentation, we consider the Pauli equation for a single particle in an external electromagnetic field following papers of Takabayasi \cite{Takabayasi PTP 55 b}-\cite{Takabayasi PTP 83}. We should also notice that the set of basic QHD equations for charged spinning particles considered in the self-consistent field approximation almost coincide with the single particle one \cite{Rand PF 64}. This coincidence has been actively used over last decade (see for instance \cite{Shukla RMP 11}, \cite{Marklund PRL07}, \cite{Brodin NJP 07}, \cite{Rand PF 64}).

Thus we start with the Pauli equation
\begin{equation}\label{SUSD Pauli} \imath\hbar\partial_{t}\psi=\biggl(\frac{(\frac{\hbar}{\imath}\nabla-\frac{q_{e}}{c}\textbf{A})^{2}}{2m}+q_{e}\varphi-\gamma_{e}\widehat{\mbox{\boldmath $\sigma$}} \textbf{B}\biggr)\psi \end{equation}
governing evolution of spinor wave function $\psi(\textbf{r},t)$. In equation (\ref{SUSD Pauli}) $\varphi=\varphi_{ext}$, $\textbf{A}=\textbf{A}_{ext}$ are the scalar and vector potentials of external electromagnetic fields, $\textbf{B}=\textbf{B}_{ext}$ is the external magnetic field, $q_{e}=-e$ is the charge of electron, $m$ is the mass of the particle under consideration, $\gamma_{e}$ is the gyromagnetic ratio, $\nabla$ is the gradient operator, $\mbox{\boldmath $\sigma$}$ is the vector of Pauli matrixes, $\hbar$ is the reduced Planck constant, $c$ is the speed of light.

Let us present the explicit form of the Pauli matrixes
\begin{equation}\label{QGGR}\begin{array}{ccc} \widehat{\sigma}_{x}=\left(\begin{array}{ccc}0& 1\\
1& 0\\
\end{array}\right),&
\widehat{\sigma}_{y}=\left(\begin{array}{ccc}0& -\imath \\
\imath & 0 \\
\end{array}\right),&
\widehat{\sigma}_{z}=\left(\begin{array}{ccc} 1& 0\\
0& -1\\
\end{array}\right).
\end{array}\end{equation}
The commutation relation for spin-1/2 matrixes is
\begin{equation}\label{QGGR comm rel} [\widehat{\sigma}^{\alpha}, \widehat{\sigma}^{\beta}]=2\imath\varepsilon^{\alpha\beta\gamma}\widehat{\sigma}^{\gamma}.\end{equation}

$\rho=\psi^{+}\psi$ is the probability density to find the particle in a point $\textbf{r}$ regardless its spin, where $\psi^{+}$ is the hermitian conjugated wave function.

The spinor wave function $\psi$ can be presented as
\begin{equation}\label{SUSD perturbations}
\psi=
\left(\begin{array}{ccc}
\psi_{\uparrow} \\
\psi_{\downarrow} \\
\end{array}\right).\end{equation}
Applying wave functions describing spin-up $\psi_{\uparrow}$ and spin-down $\psi_{\downarrow}$ states we can write probability density to find the particle in a point $\textbf{r}$ with spin-up $\rho_{\uparrow}=\mid\psi_{\uparrow}\mid^{2}$ or spin-down $\rho_{\downarrow}=\mid\psi_{\downarrow}\mid^{2}$. We also see $\rho=\rho_{\uparrow}+\rho_{\downarrow}$. Directions up $\uparrow$ (down $\downarrow$) corresponds to spins having same (opposite) direction as (to) the external magnetic field. While magnetic moments have opposite to spin directions.

In many-particle systems we have concentration of particles $n(\textbf{r},t)$, which are proportional to the probability density to find each particle in the point $\textbf{r}$, hence we have $n_{\uparrow}=\langle\rho_{\uparrow}\rangle$ and $n_{\downarrow}=\langle\rho_{\downarrow}\rangle$. Full concentration of particles in the sum of the particle concentrations with spin-up and spin-down $n=n_{\uparrow}+n_{\downarrow}$. The spin density $S_{z}$ of electrons is the difference between concentrations of electrons with different projection of spin $S_{z}=n_{\uparrow}-n_{\downarrow}$. Its definition is $S_{z}=\psi^{+}\sigma_{z}\psi$.
We have that the z-projection of the spin density $S_{z}$ is not an independent variable in this representation of the quantum hydrodynamics.

We can derive equations for $\rho_{\uparrow}$, and $\rho_{\downarrow}$. They are analogous to the continuity equations, but number of particles with different spin projection (or corresponding probability for a single particle) are not constants.

Let us rewrite the Pauli equation (\ref{SUSD Pauli}) in more explicit form
$$\imath\hbar\partial_{t}\psi_{\uparrow}=\biggl(\frac{(\frac{\hbar}{\imath}\nabla-\frac{q_{e}}{c}\textbf{A})^{2}}{2m}+q_{e}\varphi$$
\begin{equation}\label{SUSD Pauli Expl +} -\gamma_{e}B_{z}\biggr)\psi_{\uparrow}
-\gamma_{e}(B_{x}-\imath B_{y})\psi_{\downarrow}, \end{equation}
and
$$\imath\hbar\partial_{t}\psi_{\downarrow}=\biggl(\frac{(\frac{\hbar}{\imath}\nabla-\frac{q_{e}}{c}\textbf{A})^{2}}{2m}+q_{e}\varphi$$
\begin{equation}\label{SUSD Pauli Expl -} +\gamma_{e}B_{z}\biggr)\psi_{\downarrow}
-\gamma_{e}(B_{x}+\imath B_{y})\psi_{\uparrow}. \end{equation}
Directions spin-up and spin-down are related to a preferable direction in space. If we have an uniform external magnetic field its direction can be taken as preferable direction. In this case only z-projection of the magnetic field $B_{z}$ enters the Pauli equation for a single particle in the external magnetic field. However we going to apply corresponding QHD equations for plasma description, where motion of charges and spin evolution create $B_{x}$ and $B_{y}$.

Considering time evolution of the probability densities $\rho_{\uparrow}$ and $\rho_{\downarrow}$ we derive the continuity equations
\begin{equation}\label{SUSD cont eq electrons spin UP}
\partial_{t}n_{\uparrow}+\nabla(n_{\uparrow}\textbf{v}_{\uparrow})=\frac{\gamma}{\hbar}(B_{y}S_{x}-B_{x}S_{y}), \end{equation}
and
\begin{equation}\label{SUSD cont eq electrons spin DOWN}
\partial_{t}n_{\downarrow}+\nabla(n_{\downarrow}\textbf{v}_{\downarrow})=\frac{\gamma}{\hbar}(B_{x}S_{y}-B_{y}S_{x}), \end{equation}
where we have applied $S_{x}$ and $S_{y}$ for mixed combinations of $\psi_{\uparrow}$ and $\psi_{\downarrow}$. Their explicit form is presented and discussed below.

Usually the continuity equation shows conservation of the particle number. If we consider the spin-up electrons and the spin-down electrons separately, we find that particle numbers change due to interaction. The total number of electrons $N=N_{\uparrow}+N_{\downarrow}$ conserves only.

Particle current appears in the continuity equation in usual form $\textbf{j}_{s}=\frac{1}{2m}(\psi_{s}^{*}\textbf{D}\psi_{s}+c.c.)$, where $s=\uparrow$ or $\downarrow$, and $\textbf{D}=\widehat{\textbf{p}}-\frac{q_{e}}{m}\textbf{A}$. We have introduced the velocity fields $\textbf{v}_{s}$ via the particle currents $\textbf{j}_{s}\equiv n_{s}\textbf{v}_{s}$, with the following explicit form of the velocities $\textbf{v}_{s}=\frac{\hbar}{m}\nabla \phi_{s}-\frac{q_{e}}{mc}\textbf{A}$. Here we have applied the phase of wave function $\psi_{s}=a_{s}e^{\imath \phi_{s}}$.

Considering time evolution of the particle currents for each projection of spin $\textbf{j}_{\uparrow}$ and $\textbf{j}_{\downarrow}$ we can derive corresponding Euler equations
$$mn_{\uparrow}(\partial_{t}+\textbf{v}_{\uparrow}\nabla)\textbf{v}_{\uparrow}+\nabla p_{\uparrow}-\frac{\hbar^{2}}{4m}n_{\uparrow}\nabla\Biggl(\frac{\triangle n_{\uparrow}}{n_{\uparrow}}-\frac{(\nabla n_{\uparrow})^{2}}{2n_{\uparrow}^{2}}\Biggr)$$
$$=q_{e}n_{\uparrow}\biggl(\textbf{E}+\frac{1}{c}[\textbf{v}_{\uparrow},\textbf{B}]\biggr)+\gamma_{e}n_{\uparrow}\nabla B_{z}$$
\begin{equation}\label{SUSD Euler eq electrons spin UP} +\frac{\gamma_{e}}{2}(S_{x}\nabla B_{x}+S_{y}\nabla B_{y})+\frac{m\gamma_{e}}{\hbar}(\textbf{J}_{(M)x}B_{y}-\textbf{J}_{(M)y}B_{x}),\end{equation}
and
$$mn_{\downarrow}(\partial_{t}+\textbf{v}_{\downarrow}\nabla)\textbf{v}_{\downarrow}+\nabla p_{\downarrow}-\frac{\hbar^{2}}{4m}n_{\downarrow}\nabla\Biggl(\frac{\triangle n_{\downarrow}}{n_{\downarrow}}-\frac{(\nabla n_{\downarrow})^{2}}{2n_{\downarrow}^{2}}\Biggr)$$
$$=q_{e}n_{\downarrow}\biggl(\textbf{E}+\frac{1}{c}[\textbf{v}_{\downarrow},\textbf{B}]\biggr)-\gamma_{e}n_{\downarrow}\nabla B_{z}$$
\begin{equation}\label{SUSD Euler eq electrons spin DOWN}+\frac{\gamma_{e}}{2}(S_{x}\nabla B_{x}+S_{y}\nabla B_{y})+\frac{m\gamma_{e}}{\hbar}(\textbf{J}_{(M)y}B_{x}-\textbf{J}_{(M)x}B_{y}),\end{equation}
with
\begin{equation}\label{SUSD Spin current x} \textbf{J}_{(M)x}=\frac{1}{2}(\textbf{v}_{\uparrow}+\textbf{v}_{\downarrow})S_{x}-\frac{\hbar}{4m} \biggl(\frac{\nabla n_{\uparrow}}{n_{\uparrow}}+\frac{\nabla n_{\downarrow}}{n_{\downarrow}}\biggr)S_{y}, \end{equation}
and
\begin{equation}\label{SUSD Spin current y} \textbf{J}_{(M)y}= \frac{1}{2}(\textbf{v}_{\uparrow}+\textbf{v}_{\downarrow})S_{y}+\frac{\hbar}{4m}\biggl(\frac{\nabla n_{\uparrow}}{n_{\uparrow}}+\frac{\nabla n_{\downarrow}}{n_{\downarrow}}\biggr)S_{x}, \end{equation}
where $q_{e}=-e$, $\gamma_{e}=-g\frac{e\hbar}{2mc}$ is the gyromagnetic ratio for electrons, and $g=1+\alpha/(2\pi)=1.00116$, where $\alpha=1/137$ is the fine structure constant, gets into account the anomalous magnetic moment of electron. $\textbf{J}_{(M)x}$ and $\textbf{J}_{(M)y}$ are elements of the spin current tensor $J^{\alpha\beta}$.

Most of terms in the Euler equations (\ref{SUSD Euler eq electrons spin UP}) and (\ref{SUSD Euler eq electrons spin DOWN}) have traditional meaning. The first group of terms in the left-hand side of Euler equations are the substantial time derivatives of velocity fields $\textbf{v}_{\uparrow}$ and $\textbf{v}_{\downarrow}$. The second terms are the gradients of the thermal pressure. They do not appear from the single-particle Pauli equation, but we have included it assuming that the many-particle QHD gives this effect \cite{MaksimovTMP 1999}, \cite{MaksimovTMP 2001}, \cite{Andreev RPJ 07}. The next group of terms, proportional to the square of the Plank constant, are the contributions of the quantum Bohm potential.

The right-hand sides of Euler equations present interaction force fields. The first groups of terms in the right-hand side are the Lorentz forces. Since we consider two species of electrons these forces have same structure, with no explicit dependence on the spin direction. The implicit dependence is presented via subindexes of the concentration and velocity field. The second terms describe action of the z-projection of magnetic field on the magnetic moments (spins) of particles. Dependence on spin projection reveals in different signs before these terms. The third groups of terms in Euler equations contain a the part of well-known force field $\textbf{F}_{S}=M^{\beta}\nabla B^{\beta}$ describing action of the magnetic field on magnetic moments \cite{Maksimov Izv 2000}, \cite{Takabayasi PTP 55 b}. Part of this force field has been presented by previous terms $\textbf{F}_{S(z)}=\pm\gamma_{e}n_{\uparrow,\downarrow}\nabla B_{z}$. The second part of the force field $\textbf{F}_{S(x,y)}=\gamma_{e}(S_{x}\nabla B_{x}+S_{y}\nabla B_{y})$. The half of this force field enters each of the Euler equations. The last groups of terms is related to nonconservation of particle number with different spin-projection. This nonconservation gives extra mechanism for change of the momentum density revealing in the extra force fields.

Here we describe explicit form of spin density projections on $x$ and $y$ axes. We have used notations $S_{x}$ and $S_{y}$ in equations (\ref{SUSD cont eq electrons spin UP})-(\ref{SUSD Euler eq electrons spin DOWN}).   These quantities appear as follows $S_{x}=\psi^{*}\sigma_{x}\psi=\psi_{\downarrow}^{*}\psi_{\uparrow}+\psi_{\uparrow}^{*}\psi_{\downarrow}=2a_{\uparrow}a_{\downarrow}\cos\Delta \phi$, $S_{y}=\psi^{*}\sigma_{y}\psi=\imath(\psi_{\downarrow}^{*}\psi_{\uparrow}-\psi_{\uparrow}^{*}\psi_{\downarrow})=-2a_{\uparrow}a_{\downarrow}\sin\Delta \phi$, where $\Delta \phi=\phi_{\uparrow}-\phi_{\downarrow}$. $S_{x}$ and $S_{y}$ appear as mixed combinations of $\psi_{\uparrow}$ and $\psi_{\downarrow}$. These quantities do not related to different species of electrons having different spin direction. $S_{x}$ and $S_{y}$ describe simultaneous evolution of both species.

$S_{x}$ and $S_{y}$ are involved in equations (\ref{SUSD cont eq electrons spin UP})-(\ref{SUSD Euler eq electrons spin DOWN}). We need to derive equations for these quantities to get closed set of QHD equations. Differentiating explicit forms of $S_{x}$ and $S_{y}$ and applying the Pauli equation (\ref{SUSD Pauli Expl +}) and (\ref{SUSD Pauli Expl -}) for the time derivatives of the wave functions $\psi_{\uparrow}$ and $\psi_{\downarrow}$ we obtain the following equations
$$\partial_{t}S_{x}+\frac{1}{2}\nabla[S_{x}(\textbf{v}_{\uparrow}+\textbf{v}_{\downarrow})]$$
\begin{equation}\label{SUSD eq for Sx} -\frac{\hbar}{4m}\nabla\Biggl(S_{y}\biggl(\frac{\nabla n_{\uparrow}}{n_{\uparrow}}-\frac{\nabla n_{\downarrow}}{n_{\downarrow}}\biggr)\Biggr)=\frac{2\gamma_{e}}{\hbar}\biggl(B_{z}S_{y}-B_{y}(n_{\uparrow}-n_{\downarrow})\biggr),\end{equation}
and
$$\partial_{t}S_{y}+\frac{1}{2}\nabla[S_{y}(\textbf{v}_{\uparrow}+\textbf{v}_{\downarrow})]$$
\begin{equation}\label{SUSD eq for Sy} +\frac{\hbar}{4m}\nabla\Biggl(S_{x}\biggl(\frac{\nabla n_{\uparrow}}{n_{\uparrow}}-\frac{\nabla n_{\downarrow}}{n_{\downarrow}}\biggr)\Biggr)=\frac{2\gamma_{e}}{\hbar}\biggl(B_{x}(n_{\uparrow}-n_{\downarrow})-B_{z}S_{x}\biggr).\end{equation}
The first term in equation (\ref{SUSD eq for Sx}) (equation (\ref{SUSD eq for Sy})) is the time derivative of $S_{x}$ ($S_{y}$). The second terms in these equations are gradients of the spin fluxes. The third terms are quantum Bohm potential revealing the quantum part of the gradients of the spin fluxes. The right-hand side of equations (\ref{SUSD eq for Sx}) and (\ref{SUSD eq for Sy}) contains the torque caused by interaction of magnetic moments with the magnetic field. The right-hand side of these equations corresponds to traditional form. For instance let us consider the torque in equation for $S_{x}$, which is $T_{x}=\frac{2\gamma_{e}}{\hbar}(S_{y}B_{z}-S_{z}B_{y})=\frac{2\gamma_{e}}{\hbar}(S_{y}B_{z}-(n_{\uparrow}-n_{\downarrow})B_{y})$, that coincides with the right-hand side of equation (\ref{SUSD eq for Sx}).

Let us mention that $S_{x}$ and $S_{y}$ do not wear subindexes $\uparrow$ and $\downarrow$. As we can see from definitions of $S_{x}$ and $S_{y}$ they are related to both projections spin-up $\psi_{\uparrow}$ and spin-down $\psi_{\downarrow}$.
%simultaneously

Electromagnetic fields in the QHD equations presented above obey the Maxwell equations
\begin{equation}\label{SUSD div E} \nabla \textbf{E}=4\pi\biggl(en_{i}-en_{e\uparrow}-en_{e\downarrow}\biggr),\end{equation}
\begin{equation}\label{SUSD div B} \nabla \textbf{B}=0, \end{equation}
\begin{equation}\label{SUSD ror E} \nabla\times \textbf{E}=-\frac{1}{c}\partial_{t}\textbf{B},\end{equation}
and
$$\nabla\times \textbf{B}=\frac{1}{c}\partial_{t}\textbf{E}$$
\begin{equation}\label{SUSD rot B}
+\frac{4\pi}{c}\sum_{a=e,i}(q_{a}n_{a\uparrow}\textbf{v}_{a\uparrow}+q_{a}n_{a\downarrow}\textbf{v}_{a\downarrow})+4\pi\sum_{a=e,i}\nabla\times \textbf{M}_{a},\end{equation}
where $\textbf{M}_{a}=\{\gamma_{a}S_{ax}, \gamma_{a}S_{ay}, \gamma_{a}(n_{a\uparrow}-n_{a\downarrow})\}$ is the magnetization of electrons in terms of hydrodynamic variables.

\subsection{Equation of state}

We need to get a closed set of equations, so we should use an equation of state for the pressure for spin-up $p_{\uparrow}$ and spin-down $p_{\downarrow}$ electrons. We consider degenerate electrons. Hence, in non-relativistic case, we have
\begin{equation}\label{SUSD EqState partial}p_{s}=\frac{(6\pi^{2})^{2/3}}{5}\frac{\hbar^{2}}{m}n_{s}^{5/3}.\end{equation}
From this equation of state we find $\frac{\partial p_{s}}{\partial n_{s}}=\frac{(6\pi^{2})^{2/3}}{3}\frac{\hbar^{2}}{m}n_{s}^{2/3}$ giving contribution in the Euler equation via $\nabla p_{s}=\frac{\partial p_{s}}{\partial n_{s}}\nabla n_{s}$. Here we see that equations of state for spin-up electrons and spin-down electrons are different due to external magnetic field, which changes an equilibrium concentration of each species $n_{0\uparrow}\neq n_{0\downarrow}$. We have included that only one particle with a chosen spin direction can occupy one quantum state. As a consequence we have $(6\pi^{2})^{2/3}$ instead of $(3\pi^{2})^{2/3}$ appearing in the Fermi pressure. At derivation of the Fermi pressure one assumes that two particles with different spin directions could occupy a quantum state, but we now consider spin-up and spin-down electrons as different species.

We show below that difference between $p_{\uparrow}$ and $p_{\downarrow}$ due to difference of $n_{\uparrow}$ and $n_{\downarrow}$ leads to new effects in quantum plasmas. One of these effects is appearance of new wave, which we call the spin-electron acoustic wave.

Interactions of magnetosonic waves in a spin-1/2 degenerate quantum plasmas have been recently considered in Ref. \cite{Chang Li PP 14} in terms of quantum magnetohydrodynamics. Let us mention that the magnetohydrodynamics is very useful tool, where electron-ion plasmas are considered as a single liquid. Whereas we move in opposite direction developing many-liquid model for electrons. In Ref. \cite{Bychkov PP 10} the quantum magnetohydrodynamics was applied as well.

Considering quantum spin 1/2 plasmas researchers usually apply equation of state for unpolarized electrons
\begin{equation}\label{SUSD EqState unPol} p_{unpol}=\frac{(3\pi^{2})^{\frac{2}{3}}}{5}\frac{\hbar^{2}}{m}n^{\frac{5}{3}}, \end{equation}
see Refs. \cite{Landau v5 eq st}-\cite{Mushtaq PP 10}.

In papers \cite{Shahid PP 13}, \cite{Sharma PP 14} authors use other equations of state, but they give no change in the problem under consideration.

In Ref. \cite{Iqbal JPP 13} author presented an attempt to consider "two-fluid
model of electrons is being used which treats the spin-up and -down populations
relative to the magnetic field as different species", which appears to be incomplete. Moreover equation of state for unpolarized single liquid electrons was used there.

\section{Perturbation evolution}

Interest to spin contribution in properties of plasmas was appeared \cite{Maksimov Izv 2000} since many-particle quantum hydrodynamics of spin-1/2 particles had been derived in 2000 \cite{Maksimov Izv 2000}, \cite{Maksimov VestnMSU 2000}. Since when a lot of results have been obtained (see review papers \cite{Shukla PhUsp 2010}, \cite{Shukla RMP 11}, \cite{Uzdensky arxiv review 14}), but we should especially mention Refs. \cite{Andreev AtPhys 08}, \cite{Andreev IJMP 12}, \cite{Andreev VestnMSU 2007}, \cite{Misra JPP 10}, where some interesting effects were found in the linear regime of small perturbations in magnetized plasmas. It was shown that spin evolution leads to existence of new wave solution. There are two type of spin excitations in quantum plasmas propagating by means perturbations of the electric field \cite{Andreev VestnMSU 2007}, \cite{Misra JPP 10}, and by means perturbations of the magnetic field with no electric field involve in it (the quasi-magnetostatic regime) \cite{Andreev AtPhys 08},  \cite{Andreev IJMP 12}, \cite{Andreev VestnMSU 2007}.

Some recent researches reveal new linear wave solutions. Most of them are related to spin evolution \cite{Andreev IJMP 12}, \cite{Andreev VestnMSU 2007}, \cite{Misra JPP 10}. And a longitudinal solution, which is called the positron sound wave, was found in Ref. \cite{Tsintsadze EPJD 11}. In this section we present new longitudinal wave in degenerate electrons moving on background of motionless ions, which we call the spin-electron acoustic wave.

Here we consider propagation of waves parallel to external field. It includes consideration of spin-plasma waves propagating by means perturbations of the electric field \cite{Andreev VestnMSU 2007}, \cite{Misra JPP 10}.

Equilibrium condition is described by the non-zero concentrations $n_{0\uparrow}$, $n_{0\downarrow}$, $n_{0}=n_{0\uparrow}+n_{0\downarrow}$, and external magnetic field $\textbf{B}_{ext}=B_{0}\textbf{e}_{z}$. Other quantities equal to zero $\textbf{v}_{0\uparrow}=\textbf{v}_{0\downarrow}=0$, $\textbf{E}_{0}=0$, $S_{0x}=S_{0y}=0$.
Assuming that perturbations are monochromatic
\begin{equation}\label{SUSD perturbations}
\left(\begin{array}{ccc}
\delta n_{\uparrow}\\
\delta n_{\downarrow} \\
\delta \textbf{v}_{\uparrow} \\
\delta \textbf{v}_{\downarrow} \\
\delta \textbf{E}\\
\delta \textbf{B}\\
\delta S_{x}\\
\delta S_{y}\\
\end{array}\right)=
\left(\begin{array}{ccc}
N_{A\uparrow} \\
N_{A\downarrow} \\
\textbf{V}_{A\uparrow} \\
\textbf{V}_{A\downarrow} \\
\textbf{E}_{A}\\
\textbf{B}_{A}\\
S_{Ax}\\
S_{Ay}\\
\end{array}\right)e^{-\imath\omega t+\imath \textbf{k} \textbf{r}},\end{equation}
we get a set of linear algebraic equations relatively to $N_{A\uparrow}$, $N_{A\downarrow}$, $V_{A\uparrow}$, $V_{A\downarrow}$, $\textbf{E}_{A}$,
$\textbf{B}_{A}$, $S_{Ax}$, and
$S_{Ay}$. Condition of existence of nonzero solutions for amplitudes of perturbations gives us a dispersion equation.

Difference of spin-up and spin-down concentrations of electrons $\Delta n=n_{0\uparrow}-n_{0\downarrow}$ is caused  by external magnetic field. Since electrons are negative their spins get preferable direction opposite to the external magnetic field $\frac{\Delta n}{n_{0}}=\tanh\biggl(\frac{\gamma_{e}B_{0}}{T_{e}}\biggr)=-\tanh\biggl(\frac{\mid\gamma_{e}\mid B_{0}}{T_{e}}\biggr)$. Here, as always we consider temperature in units of energy, so we do not write the Boltzmann constant.

We consider plasmas in the uniform constant external magnetic field. We see that in linear approach numbers of electrons of each species conserves.

After some straightforward calculations we find the following dispersion equations for the longitudinal
\begin{equation}\label{SUSD Longit disp eq} 1-\frac{\omega_{Le\uparrow}^{2}}{\omega^{2}-u_{\uparrow}^{2}k^{2}}
-\frac{\omega_{Le\downarrow}^{2}}{\omega^{2}-u_{\downarrow}^{2}k^{2}}=0,\end{equation}
and the transverse
$$k^{2}c^{2}-\omega^{2}+\omega_{Le}^{2}\frac{\omega}{\omega\pm\mid\Omega\mid}$$
\begin{equation}\label{SUSD Transv disp eq} -4\pi\gamma k^{2}c^{2}\frac{2\gamma}{\hbar}\frac{n_{0\uparrow}-n_{0\downarrow}}{\omega\pm g\mid\Omega\mid}=0,\end{equation}
waves, where
\begin{equation}\label{SUSD Lengm freq} \omega_{Le(s)}^{2}=\frac{4\pi e^{2}n_{0s}}{m}\end{equation}
is the Langmuir frequency for species $s=\uparrow, \downarrow$ of electrons, $\omega_{Le}^{2}=\omega_{Le,\uparrow}^{2}+\omega_{Le,\downarrow}^{2}$ is the full Langmuir frequency, $u_{s}^{2}=\frac{2^{2/3}}{3}v_{Fe}^{2}+\frac{\hbar^{2}k^{2}}{4m^{2}}$.

Keeping in mind that $n_{0}=n_{0\uparrow}+n_{0\downarrow}$ and $M_{ez}=\gamma_{e}(n_{e0\uparrow}-n_{e0\downarrow})=\chi_{e}B_{0}$, where $\chi_{e}$ is the ratio between equilibrium magnetic susceptibility and magnetic permeability of electrons, we find no crucial difference between equation (\ref{SUSD Transv disp eq}) and results of usual QHD applied in Refs. \cite{Andreev IJMP 12}, \cite{Andreev VestnMSU 2007}, \cite{Misra JPP 10}. However a great difference appears for longitudinal waves presented by equation (\ref{SUSD Longit disp eq}).

Let us mention that two different signs in formula (\ref{SUSD Transv disp eq}) correspond to left- and right-circular polarized waves.

Now we focus our attention on equation (\ref{SUSD Longit disp eq}). If equilibrium concentrations approximately equal $n_{0\uparrow}\approx n_{0\downarrow}$, what is possible in small magnetic field, equation (\ref{SUSD Longit disp eq}) gives spectrum of the Langmuir waves
\begin{equation}\label{SUSD Langm disp} \omega^{2}=\omega_{Le}^{2}+\frac{1}{3}v_{Fe}^{2}k^{2}+\frac{\hbar^{2}k^{4}}{4m^{2}},\end{equation}
where $v_{Fe}=(3\pi^{2}n_{0})^{1/3}\hbar/m$ is the Fermi velocity. It has the well-known structure.

If we can not neglect difference between $n_{0\uparrow}$ and $n_{0\downarrow}$, which increases with increasing of external magnetic field we have the following dispersion equation
$$\omega^{4}-\omega^{2}[(u_{\uparrow}^{2}+u_{\downarrow}^{2})k^{2}+\omega_{Le\uparrow}^{2}+\omega_{Le\downarrow}^{2}]$$
\begin{equation}\label{SUSD} +(u_{\uparrow}^{2}\omega_{Le\downarrow}^{2}+u_{\downarrow}^{2}\omega_{Le\uparrow}^{2})k^{2}+u_{\uparrow}^{2}u_{\downarrow}^{2}k^{4}=0\end{equation}

General solution of the dispersion equation for the longitudinal waves appears as a couple of solutions
\begin{widetext}
$$\omega^{2}=\frac{1}{2}[(u_{\uparrow}^{2}+u_{\downarrow}^{2})k^{2}+\omega_{Le\uparrow}^{2}+\omega_{Le\downarrow}^{2}]$$
\begin{equation}\label{SUSD finite diff of n disp eq general} \pm\sqrt{(u_{\uparrow}^{2}-u_{\downarrow}^{2})^{2}k^{4}+(\omega_{Le\uparrow}^{2}+\omega_{Le\downarrow}^{2})^{2}
+2(u_{\uparrow}^{2}-u_{\downarrow}^{2})(\omega_{Le\uparrow}^{2}-\omega_{Le\downarrow}^{2})k^{2}}.\end{equation}
\end{widetext}

We now describe some limit cases of these formulas.

As the first step we consider limit of small magnetic fields and, consequently, we have small, but non-neglectable, difference between $n_{0\uparrow}$ and $n_{0\downarrow}$. In this limit we obtain
$$\omega_{+}^{2}=\omega_{Le}^{2}+\frac{1}{2}(u_{\uparrow}^{2}+u_{\downarrow}^{2})k^{2}$$
\begin{equation}\label{SUSD finite diff of n disp eq Small n Lang} +(u_{\uparrow}^{2}-u_{\downarrow}^{2})k^{2}\frac{(u_{\uparrow}^{2}-u_{\downarrow}^{2})k^{2}+2(\omega_{Le\uparrow}^{2}-\omega_{Le\downarrow}^{2})}{4(\omega_{Le\uparrow}^{2}+\omega_{Le\downarrow}^{2})},\end{equation}
and
$$\omega_{-}^{2}=\frac{1}{2}(u_{\uparrow}^{2}+u_{\downarrow}^{2})k^{2}$$
\begin{equation}\label{SUSD finite diff of n disp eq Small n Sound}  -(u_{\uparrow}^{2}-u_{\downarrow}^{2})k^{2}\frac{(u_{\uparrow}^{2}-u_{\downarrow}^{2})k^{2}+2(\omega_{Le\uparrow}^{2}-\omega_{Le\downarrow}^{2})}{4(\omega_{Le\uparrow}^{2}+\omega_{Le\downarrow}^{2})}.\end{equation}
$\omega_{-}$ presents a sound-like solution existing in electron gas due to different equilibrium distribution of spin-up and spin-down electrons.

Formula (\ref{SUSD finite diff of n disp eq Small n Lang}) presents the Langmuir wave dispersion. However the coefficient in front of $k^{2}$ has more complicate form instead of the usual contribution of the Fermi pressure $\frac{1}{3}v_{Fe}^{2}$. Equilibrium distribution of spinning particles being in the external magnetic field differs from the distribution in absence of the magnetic field. This difference reveals in  more complicated form of the equation of state. Suitable equation of state can be applied even in the single fluid model of electron motion \cite{Maksimov VestnMSU 2000}
$$p_{sf}=\frac{1}{2}\biggl[\frac{(6\pi^{2})^{\frac{2}{3}}}{5}\frac{\hbar^{2}}{m}\biggl(n_{(av)}+\frac{\Delta n}{2}\biggr)^{\frac{5}{3}}$$ \begin{equation}\label{SUSD eq State single Fl polarised} +\frac{(6\pi^{2})^{\frac{2}{3}}}{5}\frac{\hbar^{2}}{m}\biggl(n_{(av)}-\frac{\Delta n}{2}\biggr)^{\frac{5}{3}}\biggr]\end{equation} %sf=single fluid
However this effect was not included in Refs. \cite{Andreev IJMP 12}, \cite{Marklund PRL07}, \cite{Brodin NJP 07}, \cite{Andreev VestnMSU 2007}, \cite{Misra JPP 10}, \cite{Mushtaq PP 12}-\cite{Masood PP 11} at consideration of spectrum of magnetized plasmas of spinning particles. Now we consider spin-up and spin-down electrons separately having different equations of state for each of them. Hence it hard to miss this effect. So let us describe its contribution in spectrum of the Langmuir waves.

In small external magnetic field we can make expansion of $n_{\uparrow}$ and $n_{\downarrow}$ in series on small deviation of spin-up and spin-down concentrations from the average one $n_{(av)}\equiv n_{0}/2$, with $n_{\uparrow}=n_{(av)}-\Delta n/2$ and $n_{\downarrow}=n_{(av)}+\Delta n/2$. Thus we have more explicit form of solutions
$$\omega_{+}^{2}=\omega_{Le}^{2}+\frac{1}{3}v_{Fe}^{2}k^{2}\biggl[1-\frac{1}{9}\biggl(\frac{\Delta n}{n_{0}}\biggr)^{2}\biggr]+\frac{\hbar^{2}k^{2}}{4m^{2}}$$
\begin{equation}\label{SUSD finite diff of n disp eq Small n Lang explis}
+\biggl(\frac{\Delta n}{n_{0}}\biggr)^{2}\frac{v_{Fe}^{2}k^{2}}{9\omega_{Le}^{2}}\biggl(\frac{1}{9}v_{Fe}^{2}k^{2}+\omega_{Le}^{2}\biggr),\end{equation}
and
$$\omega_{-}^{2}=\frac{1}{3}v_{Fe}^{2}k^{2}\biggl[1-\frac{1}{9}\biggl(\frac{\Delta n}{n_{0}}\biggr)^{2}\biggr]+\frac{\hbar^{2}k^{2}}{4m^{2}}$$
\begin{equation}\label{SUSD finite diff of n disp eq Small n Sound explis} -\biggl(\frac{\Delta n}{n_{0}}\biggr)^{2}\frac{v_{Fe}^{2}k^{2}}{9\omega_{Le}^{2}}\biggl(\frac{1}{9}v_{Fe}^{2}k^{2}+\omega_{Le}^{2}\biggr).\end{equation}
In this limit the external magnetic field gives an extra term in the Langmuir wave dispersion dependence.

For the first step on the path of estimations we consider $n_{0}=10^{22}$ cm$^{-3}$, $k\sim10^{7}$ cm$^{-1}$, $\Delta n/n_{0}\sim10^{-2}$. In this case we can simplify formulas (\ref{SUSD finite diff of n disp eq Small n Lang explis}) and (\ref{SUSD finite diff of n disp eq Small n Sound explis})
\begin{equation}\label{SUSD finite diff of n disp eq Small n Lang explis Appr} \omega_{+}^{2}=\omega_{Le}^{2}+\frac{1}{3}v_{Fe}^{2}k^{2}\biggl[1+\frac{2}{9}\biggl(\frac{\Delta n}{n_{0}}\biggr)^{2}\biggr], \end{equation}
and
\begin{equation}\label{SUSD finite diff of n disp eq Small n Sound explis Appr} \omega_{-}^{2}=\frac{1}{3}v_{Fe}^{2}k^{2}\biggl[1+\frac{4}{9}\biggl(\frac{\Delta n}{n_{0}}\biggr)^{2}\biggr]. \end{equation}
At parameters under consideration we find that the shift of the Fermi pressure prevails the quantum Bohm potential. We see that dependence of dispersion on $\Delta n/n_{0}$ is quadratic at small magnetization.

Let us mention that in absence of spin we do not have dependence of the frequency on the magnetic field for the Langmuir waves propagating parallel to the external magnetic field.

Spins are highly polarized at large external magnetic fields. In this limit we can neglect concentration of spin-up electrons and consider $n_{0}\approx n_{\downarrow}$, so all spins are antiparallel to the external magnetic field. Getting into account small amount of spin-up particles we introduce the following variables $n_{\downarrow}=n_{0}-\delta$, $n_{\uparrow}=\delta$, $\Delta n=n_{0}-2\delta$, $\delta\ll n_{\downarrow}$, $\delta\ll n_{0}$, $\delta\ll\Delta n$. In this limit the general dispersion dependence (\ref{SUSD finite diff of n disp eq general}) simplifies to
$$\omega_{+}^{2}=\omega_{Le}^{2}+\frac{1}{3}2^{2/3}v_{Fe}^{2}k^{2}\biggl(1-\frac{2}{3}\frac{\delta}{n_{0}}\biggr)+\frac{\hbar^{2}k^{2}}{4m^{2}}$$
\begin{equation}\label{SUSD large diff of n Lang}  -\omega_{Le}^{2}\frac{\delta}{n_{0}} \frac{\frac{1}{3}2^{2/3}v_{Fe}^{2}k^{2}}{\omega_{Le}^{2}+\frac{1}{3}2^{2/3}v_{Fe}^{2}k^{2}},\end{equation}
and
$$\omega_{-}^{2}=\frac{1}{3}2^{2/3}v_{Fe}^{2}\biggl(\frac{\delta}{n_{0}}\biggr)^{2/3}k^{2}$$
\begin{equation}\label{SUSD large diff of n Sound} +\omega_{Le}^{2}\frac{\delta}{n_{0}} \frac{\frac{1}{3}2^{2/3}v_{Fe}^{2}k^{2}}{\omega_{Le}^{2}+\frac{1}{3}2^{2/3}v_{Fe}^{2}k^{2}}.\end{equation}
If we neglect $\Delta n/n_{0}$ in formula (\ref{SUSD large diff of n Lang}) ($\omega_{+}^{2}=\omega_{Le}^{2}+\sqrt[3]{2}\frac{1}{3}v_{Fe}^{2}k^{2}+\frac{\hbar^{2}k^{4}}{4m^{2}}$) we find the increase of the Fermi pressure contribution in $\sqrt[3]{2}$ times in compare with the Fermi pressure of unpolarized systems usually applied in literature \cite{Andreev IJMP 12}, \cite{Marklund PRL07}, \cite{Brodin NJP 07}, \cite{Andreev VestnMSU 2007}, \cite{Misra JPP 10}, \cite{Mushtaq PP 12}.

Formulas (\ref{SUSD large diff of n Lang}) and (\ref{SUSD large diff of n Sound}) are obtained for the large magnetization. Formula (\ref{SUSD large diff of n Lang}) shows linear dependence of $\omega_{+}^{2}$ on $\Delta n/n_{0}$. Spin-electron acoustic wave dispersion $\omega_{-}^{2}(k)$ has two terms with different dependence on $\Delta n/n_{0}$. One of them has linear dependence and another one proportional to $(\Delta n/n_{0})^{\frac{2}{3}}$.

The spin-electron acoustic wave is a long-frequency solution. Consequently its properties might be affected by ion motion. This problem will be considered during further development and application of the spin separated QHD model developed in this paper.

Now we move to description of equation (\ref{SUSD Transv disp eq}). The first two terms in equation (\ref{SUSD Transv disp eq}) describe propagation of the light in vacuum. The third term presents contribution of medium of charged particles moving in the external magnetic field. The last term presents medium of spinning particles. The last term exists even for neutral particles. Each of the last two terms increase degree of the dispersion equation on one in compare with the mediumless case. Simultaneous account of these two terms increase degree of the dispersion equation on two due to difference of denominators of these terms. Difference of denominators caused by the anomalous magnetic moment of electrons. If we neglect the anomalous magnetic moment of electrons we find that account of the electron spin does not change degree of dispersion equation. It gives contribution in coefficients of the equation only. Corresponding spin-plasma waves are described in Refs. \cite{Andreev IJMP 12} and \cite{Misra JPP 10}. The quantum Bohm potential in the spin evolution equation \cite{Takabayasi PTP 55 b} (see also the second term in equation (\ref{SUSD eq of magnetic moments evol}) of this paper) of the single fluid QHD model of electrons gives shift of the cyclotron frequency of magnetic moment rotation. Thus it leads to appearance of spin-plasma wave along with the along with the anomalous part of the magnetic moment \cite{Andreev Asenjo 13}, \cite{Trukhanova PrETP 13} (see also \cite{Andreev arXiv 14 positrons}).

\section{Conclusions}

We have derived QHD equations for charged spin-1/2 particles considering evolution of electrons with spin-up and spin-down separately. These equations appear as a generalization of usual quantum hydrodynamics, where physical quantities appear via contribution of all particles together, with spin-up and spin-down. This generalization reveals in existence of new wave solution and possibility to find more new solutions.

We have studied propagation of waves parallel to external magnetic field. We have found contribution of magnetic field in the Langmuir wave dispersion via difference of occupation of spin-up and spin-down states. We have considered limits of small and large magnetic field, which reveals in small and large spin polarization $\Delta n/n_{0}$ and contribution of $\Delta n/n_{0}$ in dispersion dependence. Similarly we have described new solution. It appears as a sound-like solution, which we call spin-electron acoustic wave. We have a general form of this solution and considered its limits for small and large magnetization.

%%%%%%%%%%%%%%%%%%%%%%%%%%%%%%%%%%%%%%%%%%%%%

\begin{acknowledgements}
The author thanks Professor L. S. Kuz'menkov for fruitful discussions.
\end{acknowledgements}

\end{document}